\documentclass[letterpaper]{article} 

\usepackage{aaai2026}
\usepackage{times}  
\usepackage{helvet}  
\usepackage{courier}  
\usepackage[hyphens]{url}  
\usepackage{graphicx} 
\usepackage{natbib}  
\usepackage{caption} 
\usepackage{algorithm}
\usepackage{algorithmic}

\usepackage{newfloat}
\usepackage{listings}

\usepackage{multirow}
\usepackage{multicol}
\usepackage[switch]{lineno}

\usepackage{fancybox}    
\usepackage{framed}      
\usepackage{listings}
\usepackage[tightpage]{preview}
\usepackage{colortbl}  
\usepackage{booktabs}
\usepackage{amsmath,amssymb,amsfonts}
\usepackage{amsfonts}
\usepackage{algorithmic}
\usepackage{textcomp}
\usepackage{xcolor}
\usepackage{url}

\usepackage[most]{tcolorbox}

\lstdefinelanguage{JavaScript}{
  keywords={const, let, var, if, else, for, while, function, return, null, true, false, throw, new, await, async, catch, then},
  sensitive=true,
  comment=[l]{//},
  morecomment=[s]{/*}{*/},
  string=[b]",
  morestring=[b]',
  morestring=[b]`,
}

\lstdefinestyle{javascript}{
  language=JavaScript,
  basicstyle=\ttfamily\tiny, 
  numbers=left,
  breaklines=true,
  showstringspaces=false,
  keywordstyle=\bfseries\color{blue},
  commentstyle=\itshape\color{green!50!black},
  stringstyle=\color{orange},
  framesep=5pt,
  rulecolor=\color{black},
  xleftmargin=5pt,
  xrightmargin=5pt
}

\lstdefinestyle{C}{
  language=C,
  basicstyle=\ttfamily\footnotesize,
  keywordstyle=\color{blue},
  commentstyle=\color{gray},
  stringstyle=\color{purple},
  numbers=left,
  numberstyle=\tiny\color{gray},
  stepnumber=1,
  breaklines=true,
 moredelim={[is][\textbf]{@}{@}} 
}

\lstdefinestyle{Java}{
  language=Java,
  basicstyle=\ttfamily\footnotesize,
  keywordstyle=\color{blue},
  commentstyle=\color{gray},
  stringstyle=\color{purple},
  numbers=left,
  numberstyle=\tiny\color{gray},
  stepnumber=1,
  breaklines=true,
 moredelim={[is][\textbf]{@}{@}} 
}

\newcommand{\phead}[1]{\vspace{1mm} \noindent {\bf #1}}

\definecolor{darkgreen}{rgb}{0.0, 0.5, 0.0}

\newcommand{\toolS}{Secure-Instruct\space}
\newcommand{\benchmark}{CWEBench}

\newcommand{\tool}{Secure-Instruct}

\newcommand{\toolsecure}{Secure-Instruct\textsubscript{\textit{secure}}}
\newcommand{\toolpair}{Secure-Instruct\textsubscript{\textit{(vul, secure)}}}

\newcommand{\pa}[1]{\noindent\textbf{#1}}

\definecolor{highlight}{rgb}{0.80, 0.90, 0.80}

\newtcolorbox[auto counter]{summary}[1][]{title={\bfseries },enhanced,
	coltitle=black,
	top=1mm,
	left=1mm,
	right=1mm,
	boxsep=1mm,
	attach boxed title to top left=
	{xshift=1.5em,yshift=-\tcboxedtitleheight/2},
	boxed title style={size=small,colback=lightgray},#1}

\DeclareCaptionStyle{ruled}{labelfont=normalfont,labelsep=colon,strut=off} 
\floatstyle{ruled}
\newfloat{listing}{tb}{lst}{}
\floatname{listing}{Listing}
\title{Secure-Instruct: An Automated Pipeline for Synthesizing Instruction-Tuning Datasets Using LLMs for Secure Code Generation}
\author{
    Junjie Li\textsuperscript{\rm 1}, Fazle Rabbi\textsuperscript{\rm 1}, Bo Yang\textsuperscript{\rm 1}, Song Wang,\textsuperscript{\rm 2} Jinqiu Yang\textsuperscript{\rm 1}  \\   
}
\affiliations{
    \textsuperscript{\rm 1}Concordia University\\
    \textsuperscript{\rm 2}York University\\

    \{l\_unjie, b\_yang20\}@encs.concordia.ca,
    fazle.rabbi@mail.concordia.ca, wangsong@yorku.ca, jinqiu.yang@concordia.ca
    
}

\usepackage{bibentry}

\begin{document}

\maketitle

\begin{abstract}
Although Large Language Models (LLMs) show promising solutions to automated code generation, they often produce insecure code that threatens software security. 
Current approaches (e.g., SafeCoder) to improve secure code generation are limited by small, imbalanced instruction-tuning datasets. In this work, we present~\tool{}, a novel pipeline that automatically synthesizes high‑quality vulnerable and secure code examples and instruction-tunes LLMs to align task description and secure code generation abilities.


We evaluate~\toolS on four representative LLMs using two security-related benchmarks: our own CWEBench and the existing CWEval. 
CWEBench comprises 93 scenarios on 44 CWEs, all without overlap with \tool’s synthetic instruction‑tuning dataset, while CWEval covers 31 CWEs with 119 manually verified security‑critical tasks. 
We find that \toolS improves both security and functional correctness in code generation. 
On CWEBench, \toolS substantially improves secure code generation, giving a 28.5\% increase on average in secure ratio over the pre-trained models and outperforms SafeCoder by 12.6\%.
On CWEval, \toolS achieves an increase of 157.3\% for CodeLlama-7B and 46.4\% for Mistral-7B in Func-Sec@1 over pretrained models, and significantly outperforms SafeCoder.

\end{abstract}

\section{Introduction}

Large Language Models (LLMs), trained on billions of parameters across extensive corpora, have significantly advanced various software engineering tasks, such as code generation~\citep{wei2024magicoder,luo2023wizardcoder, soen101} and program repair~\citep{jin2023inferfix,joshi2023repair}. Prominent models such as ChatGPT~\citep{openai2024chatgpt}, Llama~\citep{touvron2023llama}, and GitHub Copilot~\citep{github2024copilot} have recently gained widespread popularity for their effectiveness in these domains.  

Despite their success, LLMs often generate code with security vulnerabilities. Pearce et al.~\citep{pearce2022asleep} revealed that approximately 40\% of code snippets generated by GitHub Copilot contain vulnerabilities, and similar issues have been reported in other models~\citep{liu2024no,sandoval2023lost}. These findings suggest that, despite advances, vulnerabilities continue to pose a significant challenge in LLM-based code generation. 

Several efforts~\citep {hajipour2023codelmsec,wang2023enhancing,he2023large,he2024instruction,li2024exploratory,li2024fine} have been made to enhance the secure code generation capabilities of LLMs. Among all, SafeCoder~\citep{he2024instruction} is the state-of-the-art (SOTA), which leverages real vulnerability fixes for instruction-tuning LLMs and outperforms other approaches. Yet, techniques like SafeCoder face two major limitations. 
First, exploiting real-world vulnerabilities and their fixes is limited by scale due to their rarity. 
For instance, SafeCoder managed to collect only 465 fixing examples covering 23 different Common Weakness Enumerations (CWEs) with 6 programming languages from \textit{145 million commits}. This not only highlights the inefficiency of the approach but also its limited scalability, as even an extensive dataset contains examples for only a small subset of CWEs. 
Second, the methodology relies on a non-trivial manual inspection in the post-processing phase to balance and clean the dataset. 
Specifically, it down-samples over-represented CWE cases and manually removes examples that were incorrectly collected by the automated pipeline. Consequently, only 465 high-quality examples were retained from the 1,211 initially collected instances.

To address the abovementioned challenges, we propose \tool, \textbf{a novel and automated pipeline designed to synthesize instruction tuning datasets from descriptive documentation of CWEs}. Specifically, we utilize the CWE descriptions and code examples provided by the MITRE~\cite{mitre-cwe}
database as a foundational resource to prompt LLMs to generate diverse, high-quality, and secure code examples that LLMs can easily understand for later instruction-tuning. 
\tool{} integrates static security detection tools (SonarQube and CodeQL) to automatically ensure the high quality of LLM-synthesized datasets. 
In addition, inspired by recent works in instruction tuning~\citep{longpre2023flan}, our proposed \tool{} leverages instruction tuning to better align LLMs with secure code generation.
\tool{} (1) bridges the gap between CWE documentation and practical secure code generation, (2) improves the performance of secure code generation in LLMs, (3) requires no manual efforts in curating high-quality instruction-tuning datasets, and (4) can be easily extensible to work on new CWEs as it does not rely on open-source repositories for curating high-quality instruction-tuning dataset for new CWEs.

\tool{} experiments with two synthetic data generation schemes: (1) vulnerable-code driven and (2) secure-code driven. For (1), given a CWE description document, \tool{} prompts an LLM to generate vulnerable code first, then continuously fixes the vulnerable code to achieve secure code, resulting in code pairs of \textit{(vul, secure)}. For (2), \tool{} prompts an LLM to generate secure code practices directly \textit{(secure)}. 
Then, \tool{} leverages LLMs to generate an instruction for each code snippet. 
Last, \tool{} employs two instruction tuning strategies: one is the standard one~\cite{wei2024magicoder}, and the other is the secure instruction-tuning approach proposed by He et al.~\cite{he2024instruction}. This approach uses an instruction-tuning technique designed to learn secure coding patterns and assign penalties to vulnerable code snippets. We evaluate \tool{} on four representative LLMs (CodeLlama-7B, CodeGen2-7B, Mistral-7B, and StarCoder-1B), which are commonly used in prior works on LLM secure code generation~\citep{he2023large, he2024instruction} for fair comparison and are suitable under limited computing resources.
Our evaluation includes two benchmarks, \textit{\benchmark} (94 security scenarios covering 44 CWEs), and \textit{CWEval}~\cite{peng2025cweval} (119 manually verified security-critical tasks covering 31 CWEs). 
On \textit{\benchmark}, \toolS outperforms the state-of-the-art secure code generation approach, SafeCoder~\citep{he2024instruction} by a large margin, with an average improvement of 12.6\%. 
On \textit{CWEval}, \toolS achieves better performance with an average 56.6\% improvement of Func-Sec@1 (both secure and functionally correct) compared to pre-trained models, and it outperforms SafeCoder with an average improvement of 140.8\%.
Finally, it also improves functional correctness with +39.6\% (CodeLlama-7B), +12.6\% (Mistral-7B), and +6.5\% (CodeGen2-7B) on HumanEval~\citep{chen2021codex}.
These results highlight the effectiveness of \tool{} in advancing both secure and functional code generation.

\pa{Contributions.}
\textbf{1)} We propose \tool, a fully automated framework that synthesizes high-quality datasets from CWE documentation to instruction tune LLMs for secure code generation. \tool{} is generalizable and scalable to a diverse set of CWEs and is completely free of manual effort. \textbf{2)}  We extend existing secure benchmarks to a total of 93 security-sensitive prompts covering 44 CWEs to provide a comprehensive evaluation on secure code generation. \textbf{3)}  We conduct a comprehensive evaluation for \toolS on four LLMs, and we take a comparative analysis of secure performance with the SOTA approach. Our approach not only significantly improves the secure code generation but also improves the functionality correctness evaluated on CWEval, HumanEval, and MBPP benchmarks. \textbf{4)}  We share the synthesized instruction-tuning dataset, the replication code, and models' weights at \url{https://zenodo.org/records/16715355}.


\section{Background and Related Work}
\label{sec:related_work}

\pa{Security Concerns of LLM-Generated Code.}
While LLMs show promise in automating code generation, their training on unsanitized open-source data often leads to insecure code. Figure~\ref{fig:example_vul_gen} shows an example of vulnerable code generated by CodeLlama-7B. This vulnerable code (line 8) demonstrates a CWE-78 (`OS Command Injection'). `username' comes directly from `argv[1]', which is not sanitized or validated. An attacker could inject arbitrary commands by passing a specially crafted username as input. 

Several studies~\citep{sandoval2023lost,siddiq2023generate,gong2024well,tihanyi2025secure,fu2023security,mohsin2024can} have investigated the security vulnerabilities present in LLM-generated code. Pearce et al.~\citep{pearce2022asleep} evaluated GitHub Copilot, which is built on the OpenAI Codex model, and reported that approximately 40\% of the generated code contained security vulnerabilities. These vulnerabilities are systematically classified under the Common Weakness Enumeration (CWE)~\citep{CWE}, a widely adopted category system for security vulnerabilities. Khoury et al.~\citep{khoury2023secure} examined ChatGPT in security-critical coding scenarios, discovering that the model produced insecure code in 16 cases, with only 7 cases corrected after iterative interaction with the model. Similarly, Perry et al.~\citep{perry2023users} observed that developers relying on AI-assisted tools are more likely to introduce security vulnerabilities. 



\begin{figure}
    \centering
    \begin{lstlisting}[style=C]
int main(int argc, char *argv[]) {
    // Get the username from the command line
    char *username = argv[1];
    // Get the user info from the system
    char command[100] = {0};
    // Execute the command using the username as input
    // ** The code generated below **
    @sprintf(command, "id %s", username);@
}
    \end{lstlisting}
    \caption{An example of vulnerable code generation. Line 8 is generated by CodeLlama-7B.}
    \label{fig:example_vul_gen}
\end{figure}

\begin{figure*}[t!]
    \centering
    \includegraphics[width=\textwidth]{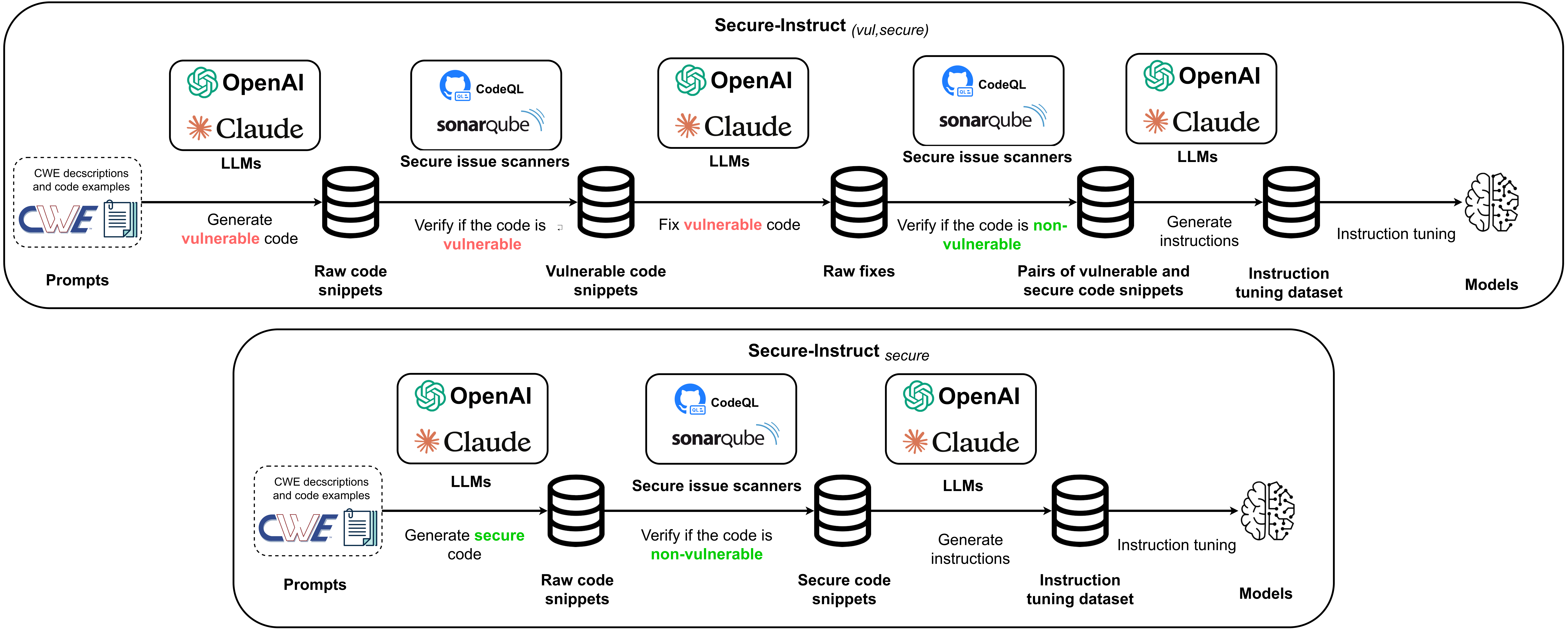} 
    \caption{An overview of \tool.
    }
    \label{fig:overview}
\end{figure*}

\pa{Improving the Security of LLM Generated Code.}
To overcome these limitations, Li et al.~\cite{li2024fine} proposed an approach to fine-tune one LLM by using the vulnerability-fixing commits. Hajipour et al.~\citep{hajipour2024hexacoder} introduced a technique to synthesize secure code examples guided by CodeQL~\citep{codeql} oracle. Another work~\citep{li2024exploratory} explored the effect of parameter-efficient fine-tuning techniques in secure code generation. He et al.~\citep{he2023large} introduced SVEN, an advanced prompting method, i.e., prefix-tuning, that uses property-specific continuous vectors (prefixes) to guide LLMs to generate secure code and reduce the likelihood of generating vulnerable code. The training of SVEN optimizes these continuous vectors using a carefully curated, high-quality dataset. 

Most recently, SafeCoder~\citep{he2024instruction} applies instruction-tuning on a dataset of 465 vulnerable-fix pairs, achieving state-of-the-art results in secure code generation.
Despite improved results, SafeCoder has several limitations. First, the quality of code from open-source repositories is inconsistent, often involving complex dependencies with external libraries. As a result, many code snippets are unsuitable for fine-tuning, requiring significant manual effort to filter and curate appropriate samples. Second, it is challenging to obtain a sufficiently large fine-tuning dataset for each category of security issues (e.g., CWE). For example, the work~\citep{he2024instruction} identified 71 code snippets for CWE-079 but only one for CWE-732, resulting in an imbalance that hinders the model's ability to achieve robust security performance. Besides, this approach is expensive. They only collected 456 code snippets covering 23 CWEs from 145 million commits, making it difficult to expand to other CWEs.
To address these limitations, we propose \tool{} to automatically construct a more extensive and diverse dataset to further enhance the security of LLM-generated code.

\section{The Methodology of \tool}
\label{sec:method}



We introduce \tool{} that leverages LLMs (e.g., ChatGPT and Claude) to synthesize data for instruction-tuning code language models to enhance secure code generation. 
Figure~\ref{fig:overview} shows an overview of \tool. 
It adopts two data-generation schemes, secure-code driven (\toolsecure) and vulnerable-code driven (\toolpair), to generate high-quality secure training data with no human intervention. 
It then uses LLMs to generate natural-language instructions that describe the functionality of each code example in the dataset. Finally, \toolS applies two instruction-tuning strategies to produce security-enhanced LLMs (Section~\ref{sec:instruct_tuning}): standard instruction tuning~\citep{longpre2023flan} and security-specialized instruction tuning~\citep{he2024instruction}. 





\subsection{CWE Seeds Collection}  
\begin{figure}
    \centering
    \includegraphics[width=0.45\textwidth]{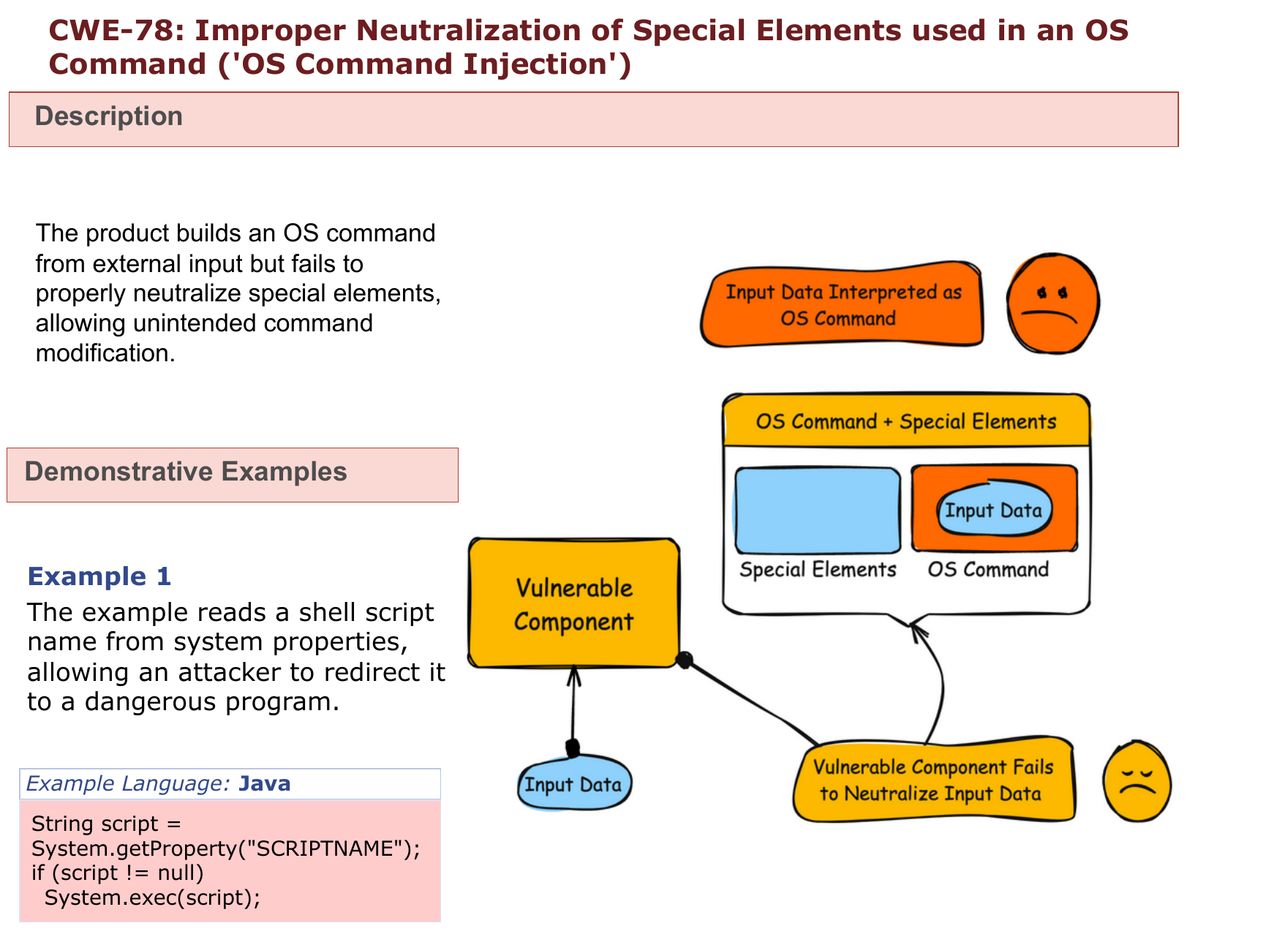}
    \caption{An example of CWE-78 on the MITRE website.}
    \label{fig:mitre}
\end{figure}

First, we collect Common Weakness Enumeration (CWE) documents from the MITRE. The MITRE is an official website that serves as an authoritative resource for understanding and analyzing CWEs. Each CWE entry on the website includes a comprehensive description, highlighting the nature of the vulnerability, its potential impact, and common scenarios in which it arises. Additionally, the entries feature illustrative code examples that contrast vulnerable and secure implementations, offering both context and practical insights. As shown in Figure~\ref{fig:mitre}, CWE-78 (OS Command Injection) provides such insights. \toolS leverages the information from the MITRE website for each CWE, namely 1) the category of CWE, 2) the overall description, and 3) code snippet examples with their explanations. In total, we collected 44 CWEs from MITRE.


\subsection{Synthetic Data Generation By Prompting LLMs}
Given seed CWE documents, \toolS employs two data generation strategies to prompt LLM for generating instruction-tuning datasets, namely vulnerable-code driven (\toolpair) and secure-code driven (\toolsecure).

\pa{\toolpair}.  As Figure~\ref{fig:overview} illustrates, the vulnerable-code driven data generation includes four steps: (1) prompting an LLM (e.g., GPT 4o) to generate vulnerable code given one CWE information; (2) verifying 
 that the LLM-generated code indeed contains the specified CWE; (3) prompting LLMs to fix the vulnerabilities and produce secure code; and (4) confirming the secure code is vulnerability-free. Vulnerable-code driven generation ensures the success of generating both vulnerable and secure code for specific CWEs.
 We employ two powerful LLMs for code generation, i.e., ChatGPT (gpt-4o-2024-08-06) and Claude (claude-3-5-sonnet-20241022), to generate intentionally vulnerable code snippets. The templates for generating and fixing vulnerable code are included in our replication package. 
For each pair of ``CWE'' and ``programming language'',  \toolpair{} takes both the CWE description and a randomly selected code example from the seed corpus and prompts each LLM to generate 10 vulnerable code snippets. 
In total, \toolpair{} collected 1,560 code snippets spanning 78 CWE-language pairs. Then, \toolpair{} utilizes static security analyzers to verify the vulnerabilities present in the generated code. 
Finally, we instruct the LLMs to fix the vulnerable code and verify that the vulnerabilities are gone using static security analyzers. Since static analysis tools can produce both false positives and false negatives, we use two static security analyzers, CodeQL~\citep{codeql} and SonarQube~\citep{sonarqube}, to complement each other in verification. 
A code snippet is considered vulnerable only if both tools produce identical detection results. 

After verification of the 1,560 snippets by static analyzers, 475 were confirmed to be truly vulnerable. These verified snippets then served as input seeds for the LLMs to generate five fixes for each snippet. Finally, static analyzers are used again to verify whether the generated fixes were valid. After generating the dataset through our pipeline and verifying it using static analyzers, we performed a de-duplication step to eliminate potential overlaps with evaluation benchmarks. This ensures that our evaluation remains fair and unbiased. The details of this de-duplication process are described in Section~\ref{sec:evaluation}.
In the end, \toolpair{} produced 856 verified (vulnerable, fixed) pairs across 25 CWEs and 6 languages, covering 39 CWE-language combinations.

\label{sec:tool_secure}
\pa{\toolsecure}. \toolpair{}'s double verification process results in many LLM-generated code snippets being filtered out, e.g., LLMs may fail to generate vulnerable code or fix vulnerable code in many cases, leading to a limited set of secure code snippets. 
To source an even larger set of secure code snippets, we propose an alternative approach, \toolsecure. Instead of generating vulnerable code and then repairing it, we directly instruct LLMs to produce secure code using CWE documents as prompts. 
For each CWE-language pair, we generate 100 secure code snippets that pass verification by both CodeQL and SonarQube, ensuring they do not contain vulnerabilities. Similarly, we performed a de-duplication step to make sure that all generated code snippets are not duplicated with evaluation benchmarks.

\subsection{Instruction Tuning LLMs for Secure Code Generation}
\label{sec:instruct_tuning}

\toolS utilizes LLMs to generate a problem description as an ``instruction'' for each of the instruction-tuning datasets. 
Then \toolS employs two instruction tuning methods, i.e., 1). Standard instruction tuning and 2). Secure instruction tuning.

\pa{Generating Instruction.}  
We follow the work~\citep{he2024instruction} to construct the instruction for each code snippet using LLMs. After that, we format the instruction-tuning datasets by putting the code snippet in the $<$response$>$ tag and the generated instructions in the $<$instruction$>$ tag, following a prior work~\citep{alpaca}.



\pa{Standard Instruction Tuning.} This process fine-tunes an LLM on instruction-response pairs to improve its ability to follow natural language instructions. It uses supervised fine-tuning (SFT), where the model learns by minimizing loss over the provided dataset, which has been adopted by many prior studies~\citep{wei2024magicoder, xu2024wizardlm, luo2023wizardcoder}.


\pa{Secure-Specialized Instruction Tuning.}
SafeCoder~\citep{he2024instruction} proposed a specialized instruction-tuning technique for secure code generation by leveraging pairs of vulnerable code and fixed code mined from open-sourced software repositories. Since we collected 856 pairs of vulnerable and from \toolpair, we applied this instruction tuning approach to our dataset. Prior research~\citep{he2023large} highlights a critical concern: exclusively using secure code examples for instruction-tuning can lead to a decline in code functionality. To address this limitation, we follow the work~\citep{he2024instruction} via the integration of a secure code dataset and a functional code dataset during instruction-tuning. It uses standard instruction tuning for the functional dataset and secure instruction tuning for the secure dataset to improve both functional and secure performance for LLMs. Specifically, we incorporate the functional code dataset from the Evol-Instruct dataset~\citep{luo2023wizardcoder}, a high-quality dataset designed to enhance the functional robustness and generalization of code generation models. 



\section{Evaluation}
\label{sec:evaluation}

\subsection{Evaluation Setup}
\label{sec:setup}

\phead{Subject Large Language Models.}
 We evaluate \toolS to improve four representative small and medium-sized LLMs with 1 to 7 billion parameters: CodeGen2-7B~\citep{nijkamp2023codegen2}, CodeLlama-7B~\citep{roziere2023code}, Mistral-7B~\citep{jiang2023mistral}, and StarCoder-1B~\citep{li2023starcoder}. These models are specifically designed for code generation, offering a balance between computational efficiency and performance. These models are widely used in prior works on secure code generation, such as \citep{he2023large,he2024instruction,wei2024magicoder} for a fair comparison. For each of the four studied LLMs, we experimented with two settings of \tool: (1) applying security-specialized instruction tuning on the dataset of pairs of vulnerable and secure code, i.e.,  from \toolpair; and (2) applying standard instruction tuning on the secure code snippets from~\toolsecure. Table~\ref{tab:fine_tuning_datasets} summarizes the statistics of the resulting fine-tuning datasets using \tool.   

\begin{table}
\centering
{\small
\setlength{\tabcolsep}{1mm}  

\begin{tabular}{lcc}
\toprule
\textbf{Attribute} & \textbf{Secure-Instruct} & \textbf{\textbf{Secure-Instruct}} \\ 
&\bf (vul, secure) &\bf secure \\
\midrule
\# of Code Snippets & 856  & 15,600 \\ \hline
\# of Average LOCs & 52.0 & 50.0 \\ \hline
\# of CWEs & 25 & 44 \\ \hline
\# of CWE-language pairs & 39 & 78 \\ \bottomrule
\end{tabular}
\caption{The statistics of the fine-tuning datasets used in \tool.}
\label{tab:fine_tuning_datasets}
}
\end{table}





\phead{Details on Instruction-Tuning LLMs.}
We use the parameter-efficient technique, LoRA~\cite{hu2021lora}, to instruction-tune the 7B LLMs. LoRA reduces the GPU memory usage and maintains a promising performance. For the 1B model, we apply full-weight instruction-tuning. All experiments are conducted on five Nvidia RTX A6000 GPUs with 48GB of memory. For full-weight instruction-tuning, we set the learning rate to 2e-5, while for LoRA fine-tuning, the learning rate is 3e-4. Each model is tuned in 2 epochs. In LoRA, we configure the LoRA alpha to 16 to control the scaling of LoRA layers and set the dropout rate to 0.05.

\phead{Evaluation Benchmarks.}
We evaluate \toolS using two secure-code generation benchmarks: (1) \textit{\benchmark{}} crafted and extended by us, and (2) \textit{CWEval}~\citep{peng2025cweval}, which enables validation of both functional correctness and security by test cases.  

First, we craft \textit{\benchmark{}} by extending prior benchmarks for secure code generation~\citep{siddiq2022securityeval, pearce2022asleep, he2024instruction}.
Such prior benchmarks are composed of security-specific scenarios and are sourced from three resources, i.e., MITRE documentation~\citep{CWE}, CodeQL repository~\citep{codeql}, and scenarios manually crafted by the authors. 
We first combined all three existing benchmarks~\citep{siddiq2022securityeval, pearce2022asleep, he2024instruction}, removed duplicates, and excluded the scenarios derived from MITRE examples, as \toolS uses MITRE documents as seeds for LLMs to generate fine-tuning datasets. We define a \textbf{secure ratio} to measure LLM's capability in CWEBench. \textit{Secure ratio} is the percentage of secure code identified by CodeQL among all the generated code. To ensure robust results, we generated 100 samples for each security-specific scenario with a temperature of 0.8. A piece of LLM-generated code is deemed \textit{secure code} if CodeQL detects no vulnerabilities in the code. 
Then, we followed a similar process as prior work~\cite{he2024instruction} to design 36 new security-specific scenarios covering 27 CWEs by adapting examples from the CodeQL repository.
In the end, our crafted \textit{CWEBench} consists of 93 security-specific scenarios covering 44 CWEs in six languages.

Second, we include a recent benchmark \textit{CWEval}~\citep{peng2025cweval}, which uses test cases for automated verification of both \textit{functional correctness} and \textit{security} without relying on static security checkers. In addition, \textit{CWEval} provides manually crafted comprehensive specifications that describe complex user intentions and logic, which means it is a much closer step to real-world applications. 
In total, \textit{CWEval} consists of 119 security scenarios covering 31 CWE types in five programming languages. \textit{CWEval} comes with three metrics, namely \textbf{Func@k}, \textbf{Sec@k} and \textbf{Func-Sec@k} to evaluate the security and functionality of generated code. In this study, we set k as 1. Specifically, \textit{Func@1} measures the percentage of security scenarios whose top-1 generated code is functionally correct (i.e., passes all functional test cases). \textit{Sec@1} is defined as the percentage of security scenarios whose top-1 solution is secure (i.e., passes all security test cases). \textit{Func-Sec@1} concerns both functional correctness and security.

\noindent\textbf{De-duplication between fine-tuning datasets and evaluation benchmarks.}
To prevent data leakage between the fine-tuning dataset and the two benchmark datasets, we performed de-duplication between training and evaluation following ~\cite{zhou2025lessleak}.
Specifically, we used MinHash with Locality-Sensitive Hashing (LSH) to detect potential duplicates. Code snippets were grouped based on similarity, and pairs with a Jaccard similarity above 0.7 were flagged as potential duplicates. After that, we removed these flagged snippets from the synthesized dataset.





\subsection{Evaluation Results} \label{sec:RQ_results}
In this section, we present our evaluation results in both secure and functional benchmarks.

\noindent\textbf{Results of CWEBench.} We evaluated the pre-trained models and instruction-tuned models by \tool{}. We also conducted a comparative analysis between \tool{} and SafeCoder~\citep{he2024instruction}.  
In particular, we reused the two released models, i.e., CodeLlma-7B and Mistral-7B, which were already instruction-tuned by SafeCoder. 
Table~\ref{tab:rq1_pretrianed_finetuned} presents the results on \textit{CWEBench}. The highest secure ratio is highlighted in bold. Across all models, \toolpair{} and \toolsecure{} demonstrate significant improvements in secure code generation. Compared with the pre-trained models, \toolS achieved the highest secure ratio across all models. 
For example, CodeLlama‑7B achieved the highest secure ratio among the four models when instruction‑tuned with \toolsecure{}, improving by 46.6\% (the secure ratio from 47.6\% to 69.8\%), compared to an increase 35.3\% under \toolpair{}.
The secure ratio of CodeGen2-7B is improved from 48.8\% to 57.0\% of secure ratio by \toolsecure{} and to 49.0\% by \toolpair. Mistral-7B achieved an increase of 28.9\% by \toolsecure. StarCoder-1B also achieves a notable improvement, i.e., secure ratio rising from 55.4\% to 67.5\%. 
Across the four LLMs, \toolS improves them to generate more secure code, and \toolsecure{} shows consistently better results than \toolpair.

Compared to SafeCoder, our two models consistently outperform the SafeCoder models, with +15.2\% in CodeLlama-7B and +9.9\% in Mistral-7B using \toolsecure. At the CWE category level, CodeLlama‑7B instruction tuned with \toolsecure{} achieves the highest or ties for highest secure ratio in 29 of 44 CWEs, outperforming both its pre-trained baseline and the SafeCoder-tuned variant. Similarly, Mistral‑7B with \toolsecure{} leads in 25 of 44 CWEs relative to its pretrained and SafeCoder‑tuned counterpart. Due to the limited space, we put all detailed analyses in our replication package.
\textbf{\textit{These results demonstrate that \toolsecure~significantly enhances secure code generation, outperforming the SOTA approach, SafeCoder}}.

\begin{table}
\centering
{\small
\setlength{\tabcolsep}{1mm}
\begin{tabular}{ccccc}
\toprule

& \multirow{3}{*}{\bf \shortstack{Code\\Llama-\\7B}} & \multirow{3}{*}{\bf \shortstack{Code\\Gen2-\\7B}} & \multirow{3}{*}{\bf \shortstack{Mistral-\\7B}} &\multirow{3}{*}{\bf \shortstack{Star\\Coder-\\1B}} \\ 
&\\
&\\
\midrule

\multirow{1}{*}{\shortstack{Pre-trained}}
 & \multirow{1}{*}{47.6\%} & \multirow{1}{*}{48.8\%} & \multirow{1}{*}{51.5\%} &  \multirow{1}{*}{55.4\%}\\  \midrule

 \multirow{2}{*}{\shortstack{SafeCoder}}
 & \multirow{2}{*}{\shortstack{60.6\%\\($\uparrow$ 27.3\%)}}  & \multirow{2}{*}{NA} & \multirow{2}{*}{\shortstack{60.4\%\\($\uparrow$ 17.3\%)}} &  \multirow{2}{*}{NA}\\ 
  & & & &
 \\ \midrule

\multirow{2}{*}{\bf \shortstack{Secure-Instruct\\(vul,secure)}} 

 & \multirow{2}{*}{\shortstack{64.4\%\\($\uparrow$ 35.3\%)}} & \multirow{2}{*}{\shortstack{49.0\%\\($\uparrow$ 0.4\%)}} & \multirow{2}{*}{\shortstack{61.3\%\\($\uparrow$ 19.0\%)}} &  \multirow{2}{*}{\shortstack{63.4\%\\($\uparrow$ 14.4\%)}}\\ 
    & & & &
 \\
 \midrule

\multirow{2}{*}{\bf \shortstack{Secure-Instruct\\secure} }
 & \multirow{2}{*}{\textbf{\shortstack{69.8\%\\($\uparrow$ 46.6\%)}}} & \multirow{2}{*}{\textbf{\shortstack{57.0\%\\($\uparrow$ 16.8\%)}}} & \multirow{2}{*}{\textbf{\shortstack{66.4\%\\($\uparrow$ 28.9\%)}}} &  \multirow{2}{*}{\textbf{\shortstack{67.5\%\\($\uparrow$ 21.8\%)}}}\\  
 & & & &
 \\
 \bottomrule

\end{tabular}
\caption{The secure ratio for pre-trained and instruction-tuned models using \toolS and SafeCoder on CWEBench. $\uparrow$ and $\downarrow$ indicate relatively change from pre-trained models. NA means the model not released by SafeCoder.}
\label{tab:rq1_pretrianed_finetuned}
}
\end{table}

\noindent\textbf{Results of CWEval.} Table~\ref{tab:rq2_cweval} presents the results on the benchmark \textit{CWEval}. 
Similar to \textit{CWEBench}, \toolS achieves the highest \textit{Func-Sec@1} across all evaluated LLMs. Notably, CodeLlama-7B and Mistral-7B show significant improvements, with increases of 157.3\% and 46.4\%, compared to their pre-trained versions. 
For \textit{Sec@1}, CodeLlama-7B, CodeGen2-7B, and StarCoder-1B achieved the highest secure ratio using \tool, with 33.7\%, 27.9\%, and 16.4\%, respectively. The results of \textit{Func@1} are put in our replication package due to limited space.

Interestingly, SafeCoder exhibits a performance drop on CWEval. 
For example, SafeCoder’s CodeLlama-7B achieves only 7.1\%  \textit{Func-Sec@1}, which is lower than the pre-trained model.
A similar decline is observed in SafeCoder’s Mistral-7B. 
These findings are consistent with the observations reported in the study~\cite{peng2025cweval}. 
\textbf{\textit{Overall, LLMs with \toolS consistently outperform both their pre-trained versions and SafeCoder models in secure code generation across CWEBench and CWEval.}}

\begin{table*}
\centering
{\small
\setlength{\tabcolsep}{1mm}

\begin{tabular}{c|cc|cc|cc|cc}
\toprule

& \multicolumn{2}{c|}{\bf \shortstack{CodeLlama-7B}} & \multicolumn{2}{c|}{\bf \shortstack{CodeGen2-7B}} & \multicolumn{2}{c|}{\bf \shortstack{Mistral-7B}} & \multicolumn{2}{c}{\bf \shortstack{StarCoder-1B}} \\ 
 & \shortstack{Sec@1} & \shortstack{Func-Sec@1}  & \shortstack{Sec@1} & \shortstack{Func-Sec@1}  & \shortstack{Sec@1} & \shortstack{Func-Sec@1}  & \shortstack{Sec@1} & \shortstack{Func-Sec@1} \\
\midrule

\multirow{1}{*}{\shortstack{Pre-trained}}  & \multirow{1}{*}{17.8\%} & \multirow{1}{*}{8.9\%}  & \multirow{1}{*}{\bf 17.9\%} & \multirow{1}{*}{7.4\%}  & \multirow{1}{*}{20.4\%} & \multirow{1}{*}{12.5\%}  & \multirow{1}{*}{16.2\%} & \multirow{1}{*}{6.1\%} \\
\hline
\multirow{2}{*}{\shortstack{ SafeCoder}} & 
\multirow{2}{*}{\shortstack{17.4\%\\($\downarrow$ 2.2\%)}} & 
\multirow{2}{*}{\shortstack{7.1\%\\($\downarrow$ 20.2\%)}} 
 & \multirow{2}{*}{NA} & \multirow{2}{*}{NA} &  
\multirow{2}{*}{\shortstack{17.4\%\\($\downarrow$ 14.7\%)}} & 
\multirow{2}{*}{\shortstack{11.5\%\\($\downarrow$ 8.0\%)}}  & \multirow{2}{*}{NA} & \multirow{2}{*}{NA}\\
&&&&&&&& \\\hline

\multirow{2}{*}{\bf \shortstack{Secure-Instruct\\(vul,secure)}} &  
\multirow{2}{*}{\shortstack{31.2\%\\($\uparrow$ 75.3\%)}} & 
\multirow{2}{*}{\bf \shortstack{22.9\%\\($\uparrow$ 157.3\%)}} & 
\multirow{2}{*}{\shortstack{17.7\%\\($\downarrow$ 1.1\%)}} & 
\multirow{2}{*}{\shortstack{7.2\%\\($\downarrow$ 2.7\%)}} &  
\multirow{2}{*}{\shortstack{25.5\%\\($\uparrow$ 25.0\%)}} & 
\multirow{2}{*}{\shortstack{17.9\%\\($\uparrow$ 43.2\%)}} & 
\multirow{2}{*}{\bf \shortstack{16.4\%\\($\uparrow$ 1.2\%)}} & 
\multirow{2}{*}{\bf \shortstack{7.4\%\\($\uparrow$ 21.3\%)}} \\
&&&&&&&& \\\hline

\multirow{2}{*}{\bf \shortstack{Secure-Instruct\\secure}} & 
\multirow{2}{*}{\bf \shortstack{33.7\%\\($\uparrow$ 89.3\%)}} & 
\multirow{2}{*}{\bf \shortstack{22.9\%\\($\uparrow$ 157.3\%)}} & 
\multirow{2}{*}{\shortstack{16.4\%\\($\downarrow$ 8.4\%)}} & 
\multirow{2}{*}{\bf \shortstack{7.5\%\\($\uparrow$ 1.4\%)}} & 
\multirow{2}{*}{\bf \shortstack{27.9\%\\($\uparrow$ 36.8\%)}} & 
\multirow{2}{*}{\bf \shortstack{18.3\%\\($\uparrow$ 46.4\%)}} & 
\multirow{2}{*}{\shortstack{15.5\%\\($\downarrow$ 4.3\%)}} & 
\multirow{2}{*}{\bf \shortstack{7.4\%\\($\uparrow$ 21.3\%)}} \\

&&&&&&&& \\

\bottomrule
\end{tabular}
\caption{Sec@1 and Func-Sec@1 results on CWEval for pre-trained LLMs and LLMs tuned by SafeCoder and \tool{} (ours).}
\label{tab:rq2_cweval}
}
\end{table*}

\noindent\textbf{Results of Functional Benchmarks.} 
We evaluated our approaches with functional benchmarks, i.e., HumanEval and MBPP, two widely used benchmarks for assessing code generation in Python. For each task, we generate 40 samples and assess LLMs' functionality using the \textit{Pass@k} metric, including \textit{Pass@1}.
\textit{Pass@k} is a standard metric for evaluating code generation, which represents the probability that at least one of the $k$ generated candidates for a given problem is correct. Table~\ref{tab:rq3_humaneval} presents the Pass@1 (Pass@10 is available in our replication package) on the HumanEval and MBPP benchmarks for pre-trained LLMs, instruction-tuned LLMs by SafeCoder and \tool. For CodeLlama‑7B, using \toolsecure{} achieves the highest \textit{Pass@1} on HumanEval at 35.6\%, outperforming both the pre-trained and SafeCoder versions.
For Mistral‑7B, the model using the \toolpair{} achieves the highest score on MBPP at 30.7\%. Codegen2-7B achieved the highest \textit{Pass@1} using \toolsecure, with 16.3\% on HumanEval and 22.8\% on MBPP.
Overall, most models show performance improvements on both datasets, except for StarCoder-1B.

\begin{table*}
\centering
{\small
\setlength{\tabcolsep}{1mm}

\begin{tabular}{c|cc|cc|cc|cc}
\toprule

& \multicolumn{2}{c|}{\bf \shortstack{CodeLlama-7B}} & \multicolumn{2}{c|}{\bf \shortstack{CodeGen2-7B}} & \multicolumn{2}{c|}{\bf \shortstack{Mistral-7B}} & \multicolumn{2}{c}{\bf \shortstack{StarCoder-1B}} \\ 
 & \shortstack{HumanEval} & \shortstack{MBPP}  & \shortstack{HumanEval} & \shortstack{MBPP}   & \shortstack{HumanEval} & \shortstack{MBPP}   & \shortstack{HumanEval} & \shortstack{MBPP}  \\
\midrule

\multirow{1}{*}{\shortstack{Pre-trained}}  
& \multirow{1}{*}{25.5\%} & \multirow{1}{*}{28.3\%}  
& \multirow{1}{*}{15.3\%} & \multirow{1}{*}{20.4\%}  
& \multirow{1}{*}{23.0\%} & \multirow{1}{*}{24.7\%}  
& \multirow{1}{*}{\bf 11.7\%} & \multirow{1}{*}{13.7\%} \\
\hline

\multirow{2}{*}{\shortstack{ SafeCoder}} & 
\multirow{2}{*}{\shortstack{32.0\%\\($\uparrow$ 25.5\%)}} & 
\multirow{2}{*}{\shortstack{33.7\%\\($\uparrow$ 19.1\%)}} 
& \multirow{2}{*}{NA} & \multirow{2}{*}{NA} 
& \multirow{2}{*}{\shortstack{31.3\%\\($\uparrow$ 36.1\%)}} & 
\multirow{2}{*}{\shortstack{29.3\%\\($\uparrow$ 18.6\%)}}  
& \multirow{2}{*}{NA} & \multirow{2}{*}{NA}\\
&&&&&&&& \\\hline

\multirow{2}{*}{\bf \shortstack{Secure-Instruct\\(vul,secure)}} &  
\multirow{2}{*}{\shortstack{34.7\%\\($\uparrow$ 36.1\%)}} & 
\multirow{2}{*}{\bf \shortstack{36.0\%\\($\uparrow$ 27.2\%)}} & 
\multirow{2}{*}{\shortstack{8.1\%\\($\downarrow$ 47.1\%)}} & 
\multirow{2}{*}{\shortstack{15.4\%\\($\downarrow$ 24.5\%)}} &  
\multirow{2}{*}{\bf \shortstack{31.6\%\\($\uparrow$ 37.4\%)}} & 
\multirow{2}{*}{\bf \shortstack{30.7\%\\($\uparrow$ 24.3\%)}} & 
\multirow{2}{*}{\shortstack{6.7\%\\($\downarrow$ 42.7\%)}} & 
\multirow{2}{*}{\shortstack{12.2\%\\($\downarrow$ 10.9\%)}} \\
&&&&&&&& \\\hline

\multirow{2}{*}{\bf \shortstack{Secure-Instruct\\secure}} & 
\multirow{2}{*}{\bf \shortstack{35.6\%\\($\uparrow$ 39.6\%)}} & 
\multirow{2}{*}{\shortstack{34.7\%\\($\uparrow$ 22.6\%)}} & 
\multirow{2}{*}{\bf \shortstack{ 16.3\%\\($\uparrow$ 6.5\%)}} & 
\multirow{2}{*}{\bf \shortstack{22.8\%\\($\uparrow$ 11.8\%)}} & 
\multirow{2}{*}{\shortstack{25.9\%\\($\uparrow$ 12.6\%)}} & 
\multirow{2}{*}{\shortstack{27.1\%\\($\uparrow$ 9.7\%)}} & 
\multirow{2}{*}{\shortstack{10.9\%\\($\downarrow$ 6.8\%)}} & 
\multirow{2}{*}{\bf \shortstack{17.7\%\\($\uparrow$ 29.2\%)}} \\
&&&&&&&& \\

\bottomrule
\end{tabular}
\caption{Pass@1 results on HumanEval and MBPP for pre-trained LLMs and LLMs tuned by SafeCoder-tuned and \tool.} 
\label{tab:rq3_humaneval}
}
\end{table*}

\section{Discussions}
\label{sec:discussion}
\noindent\textbf{The accuracy of CodeQL.} In CWEBench, the ground-truth of security is based on CodeQL, which is a static tool and may produce false positives. To address this concern, we randomly sampled 291 code snippets labeled as secure by CodeQL from four models and manually verified their security. All sampled snippets were confirmed to be truly secure. To further support our conclusion, we used an additional benchmark, CWEval, which evaluates security through test cases.

\noindent\textbf{The similarity between \tool's instruction-tuning datasets and our evaluation benchmark.} We performed a similarity analysis between the instruction-tuning data and benchmark data. Specifically, we first extracted all the secure code examples 
 from the CodeQL repository for the 44 CWEs of our benchmark.
We compare each LLM-generated secure code with the secure code examples from the CodeQL repository under the same CWE.
We then calculated the cosine similarity of each pair using TF-IDF embeddings~\citep{sparck1972statistical}.
Figure~\ref{sec:discussion} depicts the average cosine similarity score for the data generated by \toolsecure{}, which is 0.17, while the average score for data generated by \toolpair{} is 0.23. \toolsecure's values align with findings from previous work~\citep{wei2024magicoder} on synthesizing instruction-tuning data using LLMs, which reported similarity scores typically ranging between 0.1 and 0.2 in the previous synthesis techniques. 
Importantly, this result indicates that the improvements from \toolS are not simply due to the inclusion of data from the same distribution of the CodeQL repository. Instead, the generated data provides additional diversity for the instruction-tuning dataset.
\begin{figure}
    \centering
\includegraphics[width=0.80\linewidth]{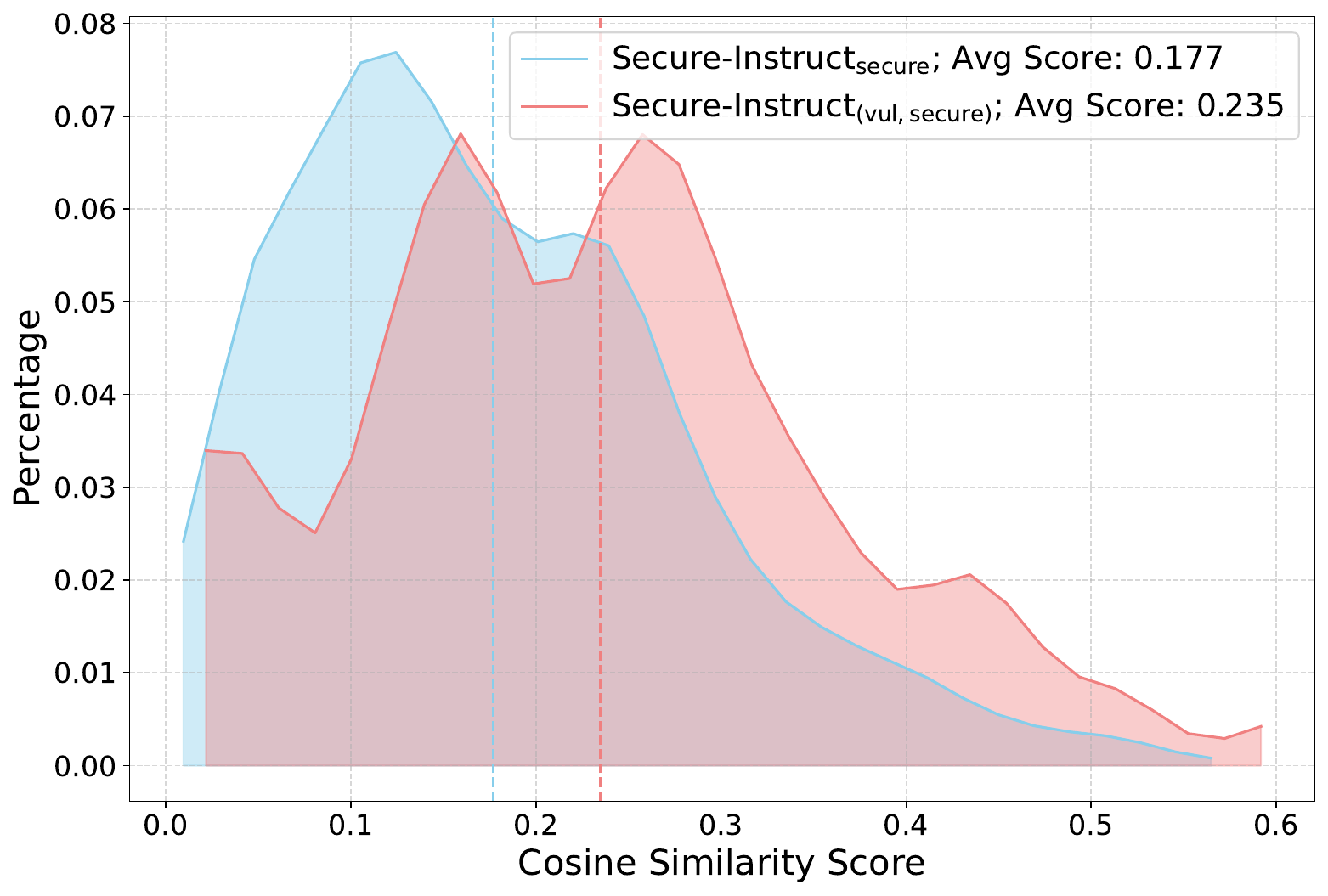} 
    \caption{Cosine similarities between \tool's instruction-tuning dataset and the secure code examples by CodeQL (i.e., our evaluation benchmark).}
    \label{fig:cosine_similarity}

\end{figure}

\noindent\textbf{Diversity of \tool's synthetic datasets.} We followed a similar procedure to compute the pair-wise similarity for \toolpair{} and \toolsecure{} using TF-IDF. The similarity of \toolpair{} is 0.17. The similarity of \toolsecure{} is 0.16. Results show that datasets from both approaches are basically diverse.  

\noindent\textbf{Low cost of \toolS in generating instruction-tuned datasets.} \tool{} generates high-quality instruction-tuning data with low cost and no human effort. Using ChatGPT (GPT-4o-2024-08-06), the total cost for generating all secure code snippets is USD 114.09.
Similarly, the cost is USD 109.74  for Claude. The cost per CWE-language is USD 2.87. 
In contrast, SafeCoder~\citep{he2024instruction} faces scalability challenges, filtering 145M commits down to just 1,200 code pairs, and requiring manual selection of 465 high-quality examples. \tool{} eliminates such overhead, offering a faster, cheaper, and automated alternative for generating secure and reliable datasets.



\begin{table}
\centering
{\small
\setlength{\tabcolsep}{1mm}
\begin{tabular}{ccccc}
\toprule

& \multirow{3}{*}{\bf \shortstack{Code\\Llama-\\7B}} & \multirow{3}{*}{\bf \shortstack{Code\\Gen2-\\7B}} & \multirow{3}{*}{\bf \shortstack{Mistral-\\7B}} &\multirow{3}{*}{\bf \shortstack{Star\\Coder-\\1B}} \\ 
&\\
&\\
\midrule

\multirow{1}{*}{\bf \shortstack{\toolpair} }
 & \multirow{1}{*}{\textbf{\shortstack{64.4\%}}} & \multirow{1}{*}{\shortstack{49.0\%}} & \multirow{1}{*}{\shortstack{61.3\%}} &  \multirow{1}{*}{\shortstack{63.4\%}}\\ 

 \midrule

\multirow{1}{*}{\bf \shortstack{\toolsecure} }
 & \multirow{1}{*}{\shortstack{61.9\%}} & \multirow{1}{*}{\textbf{\shortstack{49.1\%}}} & \multirow{1}{*}{\textbf{\shortstack{62.0\%}}} &  \multirow{1}{*}{\textbf{\shortstack{66.2\%}}}\\  

 \bottomrule

\end{tabular}
\caption{The secure ratios of \toolpair{} and \toolsecure{} with the same size (865) of training dataset.}
\label{tab:size_effect}

}
\end{table}

 \noindent\textbf{\toolpair{} v.s. \toolsecure.} \toolsecure{} uses a much larger dataset than \toolpair{} (15.6K vs. 856), so its better performance is expected. To compare fairly, we downsampled \toolsecure{} to 865 examples, covering all CWE-language pairs. 
Table~\ref{tab:size_effect} shows the comparison results across the four LLMs. 
Notably, the differences between the two are minimal, indicating essentially equivalent effectiveness. This shows that \toolsecure{} still generates highly effective synthetic data without being driven by vulnerable code first. 



\section{Conclusion}\label{sec:conclusion}
We presented \tool, a novel framework for improving secure code generation by leveraging LLMs to synthesize instruction-tuning datasets. 
\toolS overcomes the key limitations of the SOTA technique, particularly addressing dataset imbalance, inefficient data curation, and the significant manual effort required. Unlike existing approaches, \toolS can be easily adapted to support new CWEs. We also introduce CWEBench, the largest benchmark for secure code generation. Our evaluation shows that \toolS significantly improves the performance of four LLMs, achieving an average improvement of 28.5\% on CWEBench and a 140.8\% improvement of Func-Sec@1 on CWEval. Results show that \tool{} outperforms the SOTA method, demonstrating both scalability and efficiency. 


\footnotesize

\bibliography{aaai2026,ref}



\clearpage
\appendix

\section{Appendix A: Prompt Template in Data Synthesizing}  
In our study, we leverage two schemes to synthesize data by vulnerable-code driven for \toolpair{} and secure-code driven for \toolpair{}. In \toolpair, Figure~\ref{fig:prompt_template_vulnerable} shows the prompt template that we first prompt LLMs to generate vulnerable code snippets. Then another prompt is used for vulnerable code fixing, shown in Figure~\ref{fig:prompt_template_fixing}.

In \toolsecure, the prompt shown in Figure~\ref{fig:prompt_template_secure} is similar to the first step of \toolpair, and the only change is that we directly ask LLMs to generate secure code snippets.

\begin{figure}[htbp]
\centering
\begin{tcolorbox}[width=\linewidth, colback=gray!5, colframe=black!75, title=\textbf{Prompt Template for Vulnerable Code Generation of \toolpair}]
\small
\textbf{Task:} Please generate an easily understandable \textbf{vulnerable} code snippet based on the following description of CWE. (Please consider the diversity of the \textbf{vulnerable} code examples for each generation.)
\vspace{0.2cm}  

\textbf{CWE:}  
\texttt{\{<CWE\_OVERALL\_DESCRIPTION>\}}
\vspace{0.2cm}  

\textbf{A Vulnerable Code Example:}  
This is \texttt{\{<LANGUAGE>\}} language example.  
\begin{verbatim}
{<CODE>}
\end{verbatim}
\vspace{0.2cm}
\textbf{Explanation of the Example:}  
\texttt{\{<EXPLANATION>\}}
\vspace{0.2cm}  

\textbf{Can you generate a vulnerable code example for }\texttt{\{<TARGET\_LANGUAGE>\}} \textbf{language?}
\end{tcolorbox}
\caption{Prompt template used for vulnerable code generation in \toolpair.}
\label{fig:prompt_template_vulnerable}
\end{figure}

\begin{figure}[htbp]
\centering
\begin{tcolorbox}[width=\linewidth, colback=gray!5, colframe=black!75, title=\textbf{Prompt Template for Fixing Vulnerable Code of \toolpair}]
\small
\textbf{Task:} Please fix the vulnerable code based on the following description of CWE:

\vspace{0.2cm}
\textbf{CWE:}    
\texttt{\{<CWE\_OVERALL\_DESCRIPTION>\}}

\vspace{0.2cm}
\textbf{The Vulnerable Code:}    
\begin{verbatim}
{<GENERATED_VULNERABLE_CODE>}
\end{verbatim}

\vspace{0.2cm}
\textbf{Can you FIX the code for} \texttt{\{<LANGUAGE>\}} \textbf{? Please make sure the code is secure to the CWE mentioned above and runnable. }

\textbf{Task:} Please generate an easily understandable secure code snippet based on the following description of CWE. (Please consider the diversity of the vulnerable code examples for each generation.)

\end{tcolorbox}
\caption{Prompt template used for vulnerable code fixing in \toolpair.}
\label{fig:prompt_template_fixing}
\end{figure}

\begin{figure}[htbp]
\centering
\begin{tcolorbox}[width=\linewidth, colback=gray!5, colframe=black!75, title=\textbf{Prompt Template for Secure Code Generation of \toolsecure}]
\small
\textbf{Task:} Please generate an easily understandable \textbf{secure} code snippet based on the following description of CWE. (Please consider the diversity of the \textbf{secure} code examples for each generation.)
\vspace{0.2cm}  

\textbf{CWE:}  
\texttt{\{<CWE\_OVERALL\_DESCRIPTION>\}}
\vspace{0.2cm}  

\textbf{A Vulnerable Code Example:}  
This is \texttt{\{<LANGUAGE>\}} language example.  
\begin{verbatim}
{<CODE>}
\end{verbatim}
\vspace{0.2cm}
\textbf{Explanation of the Example:}  
\texttt{\{<EXPLANATION>\}}
\vspace{0.2cm}  

\textbf{Can you generate a secure code example for }\texttt{\{<TARGET\_LANGUAGE>\}} \textbf{language?}
\end{tcolorbox}
\caption{Prompt template used for secure code generation in \toolsecure.}
\label{fig:prompt_template_secure}
\end{figure}














\begin{figure}[htbp]
\centering
\begin{tcolorbox}[width=\linewidth, colback=gray!5, colframe=black!75, title=\textbf{Prompt Template for Instruction Generation}]
\small
Create a single, very short (maximum two sentences) non-detailed functionality description that could be used as a prompt to generate either of the code snippets below...

\texttt{\{<The\_Non-vulnerable\_code\_snippet>\}}

\end{tcolorbox}
\caption{Prompt template used for instruction generation.}
\label{fig:prompt_template}
\end{figure}

\section{Appendix B: Distribution of CWEBench}
\begin{figure*}
    \centering
\includegraphics[width=\linewidth]{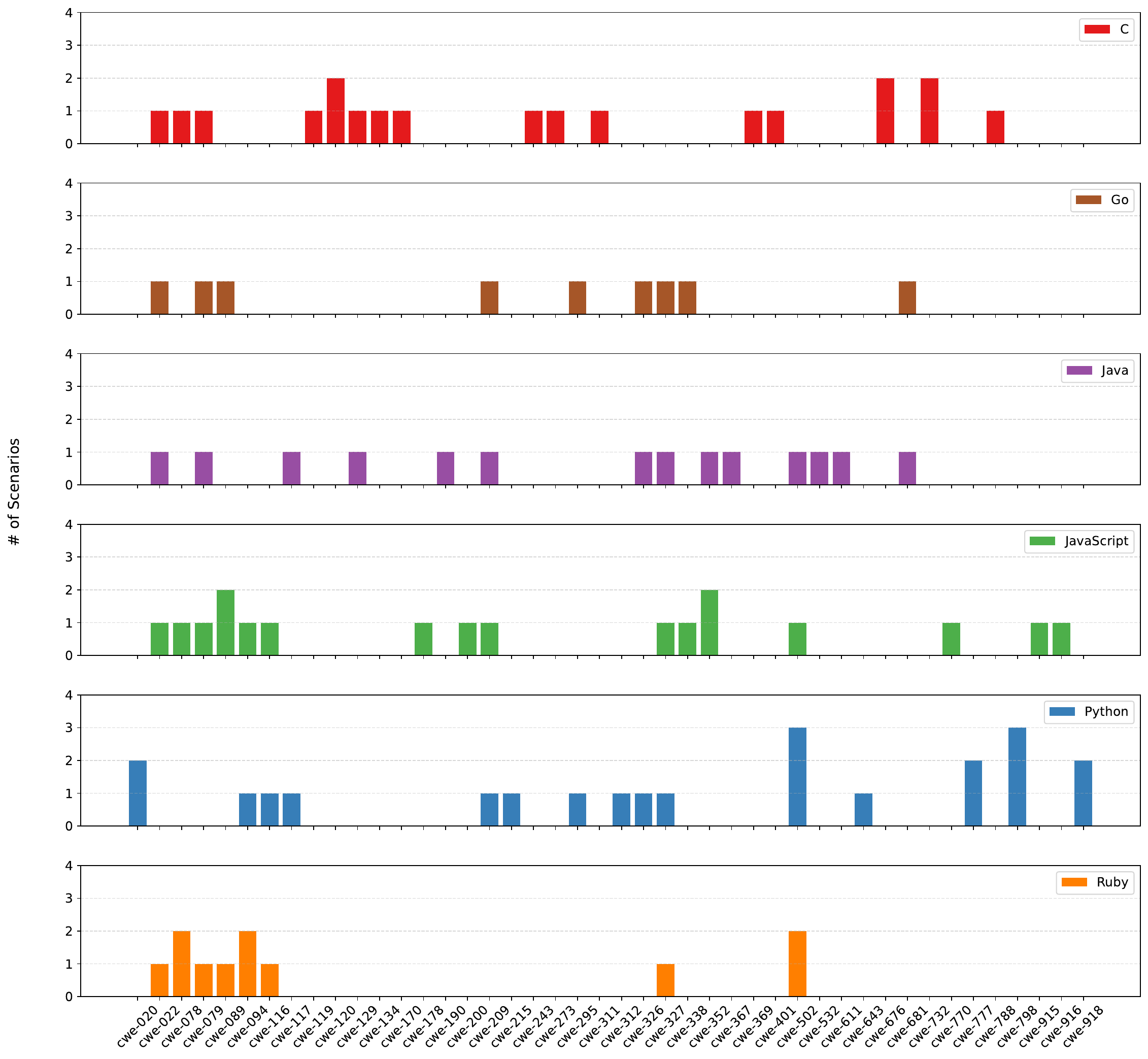} 
    \caption{The distribution of languages and CWEs in CWEBench.}
    \label{fig:cwe_language_distribution}

\end{figure*}

Figure~\ref{fig:cwe_language_distribution} shows the distribution of evaluated scenarios for languages and CWEs. Among 93 scenarios, 19 are for C, 9 are for Go, 14 are for Java, 18 are for JavaScript, 22 are for Python, and 11 are for Ruby. These scenarios cover 44 distinct CWEs, enabling a comprehensive evaluation.

\section{Appendix C: Detailed results of CWEBench}

Table~\ref{tab:rq1_codellama_results} breaks down the detailed results on CWEBench of CodeLlama-7B and Mistral-7B in their pre-trained, SafeCoder, and \tool, on each CWE in our evaluation benchmark. The highest secure ratio for each CWE is highlighted in bold.
In the scenarios whose CWEs are covered in both approaches (the top section of the table), our analysis shows that CodeLlama-7B fine-tuned with \toolsecure~achieved the highest overall secure ratio (70.6\%), outperforming SafeCoder by 2.3\% in secure ratio, Mistral-7B fine-tuned with \toolsecure~achieved 68.9\% secure ratio, with a 3.7\% higher than SafeCoder. 
In the generalization evaluation (i.e., with the full security-specific scenarios of our benchmark), \toolsecure{}demonstrates strong generalization capabilities across a broader range of CWEs for both CodeLlama-7B and Mistral-7B. Specifically, CodeLlama-7B with \toolsecure~significantly outperforms SafeCoder, achieving a 69.2\% secure ratio, while Mistral-7B achieves a 66.4\% secure ratio.

\begin{table*}
    \centering
        \caption{The results of secure ratio in CodeLlama-7B and Mistral-7B. The top section lists CWEs that are included in the fine-tuning dataset of SafeCoder, while the bottom lists CWEs that are not covered in the SafeCoder dataset.}
    {\small
\setlength{\tabcolsep}{1mm}  
    \begin{tabular}{crrrrrrrr}
    \toprule
        \multirow{2}{*}{\textbf{CWE}} &
        \multicolumn{4}{c}{\textbf{CodeLlama-7B}} &
        \multicolumn{4}{c}{\textbf{Mistral-7B}} 
        \\
        \cline{2-9}

        &  \textbf{Pre-trained} & \textbf{SafeCoder} & \textbf{\shortstack{Secure-Instruct\\Secure}} &  \textbf{\shortstack{Secure-Instruct\\(Vul, Secure)}} & \textbf{Pre-trained} & \textbf{SafeCoder} & \textbf{\shortstack{Secure-Instruct\\Secure}} &  \textbf{\shortstack{Secure-Instruct\\(Vul, Secure)}} 
        \\
    \midrule

    cwe-022 & 16.0 & \bf 43.2 & 32.0 & 37.2 & 14.8 & 34.6 & \bf 38.0 & 20.8 \\

    cwe-078 & 16.8 & \bf 79.0 & 50.7 & 40.8 & 30.5 & \bf 64.2 & 60.0 & 52.0 \\

    cwe-079 & 37.8 & \bf 84.6 & 64.8 & 67.0 & 29.2 & \bf 68.8 & 68.6 & 48.6 \\

    cwe-089 & 52.0 & 70.2 & \bf 82.5 & 72.5 & 51.7 & 60.8 & 67.0 & \bf 69.0 \\

    cwe-116 & 66.7 & 65.0 & 69.7 & \bf 75.3 & 65.7 & 68.7 & \bf 71.3 & 65.3 \\

    cwe-119 & 87.0 & \bf 99.0 & 67.0 & 47.0 & 93.0 & \bf 100.0 & 5.0 & 96.0 \\

    cwe-190 & \bf 24.0 & 7.0 & 23.0 & 23.0 & \bf 25.0 & 1.0 & 5.0 & 15.0 \\

    cwe-200 & 24.0 & \bf 99.0 & \bf 99.0 & 17.0 & 34.0 & \bf 98.0 & 73.0 & 50.0 \\

    cwe-295 & 10.5 & 78.5 & \bf 82.0 & 78.0 & 17.5 & 78.5 & 78.0 & \bf 85.0 \\

    cwe-326 & 40.0 & 33.3 & 83.0 & \bf 98.0 & 55.7 & \bf 72.7 & 59.3 & 60.3 \\

    cwe-327 & 47.8 & \bf 68.4 & 64.0 & 64.0 & 61.2 & 54.6 & \bf 77.0 & 74.8 \\

    cwe-338 & 38.5 & 42.0 & 56.0 & \bf 71.0 & 13.5 & 31.5 & \bf 50.5 & 46.0 \\

    cwe-352 & 87.7 & 93.3 & 95.7 & \bf 100.0 & 94.7 & 85.7 & \bf 97.3 & 96.3 \\

    cwe-502 & 58.3 & 79.1 & \bf 97.3 & 80.7 & 68.3 & 89.4 & \bf 93.9 & 80.0 \\

    cwe-611 & 4.0 & \bf 54.0 & 40.0 & 0.0 & 0.0 & \bf 51.0 & 40.0 & 0.0 \\

    cwe-676 & 38.0 & 48.5 & 47.5 & \bf 49.5 & 56.0 & 54.5 & \bf 93.0 & 53.5 \\

    cwe-681 & \bf 58.0 & 50.5 & 56.5 & 54.5 & 54.0 & 35.5 & \bf 80.5 & 50.5 \\

    cwe-732 & 32.0 & 75.0 & \bf 99.5 & 91.0 & 57.0 & 89.5 & \bf 93.5 & 92.5 \\

    cwe-915 & 24.0 & 82.0 & 95.0 & \bf 97.0 & 25.0 & 51.0 & 37.0 & \bf 77.0 \\

    cwe-916 & 94.0 & 97.0 & \bf 100.0 & 98.0 & 96.0 & 95.0 & 99.0 & \bf 100.0 \\
\midrule \shortstack{Total\\(above)}
 & 43.4 & 67.9 & \bf 70.4 & 66.3 & 48.2 & 64.9 & \bf 69.4 & 62.3 \\
\midrule
    cwe-020 & \bf 75.0 & 50.5 & \bf 75.0 & 61.0 & 61.0 & 75.0 & \bf 78.0 & 54.0 \\

    cwe-094 & 88.0 & 84.2 & \bf 99.2 & 98.0 & 88.8 & 82.0 & 91.5 & \bf 97.2 \\

    cwe-117 & \bf 14.5 & 7.5 & 9.5 & 2.5 & \bf 8.5 & 4.0 & 7.5 & 4.5 \\

    cwe-120 & 77.0 & 47.5 & \bf 94.0 & 93.5 & 58.0 & 80.5 & \bf 97.0 & 92.5 \\

    cwe-129 & 25.0 & 48.5 & \bf 90.0 & 75.0 & 26.0 & 64.0 & \bf 73.5 & 61.5 \\

    cwe-134 & \bf 27.0 & 0.0 & 3.0 & 0.0 & \bf 91.0 & 1.0 & 1.0 & 1.0 \\

    cwe-170 & 59.0 & 14.0 & \bf 72.0 & 48.0 & \bf 68.0 & 27.0 & 18.0 & 32.0 \\

    cwe-178 & 77.0 & 80.0 & 85.0 & \bf 99.0 & 93.0 & 70.0 & 93.0 & \bf 99.0 \\

    cwe-209 & 60.2 & 47.8 & \bf 77.8 & 62.3 & 74.0 & 46.8 & \bf 78.2 & 60.2 \\

    cwe-215 & 42.0 & 31.0 & 62.0 & \bf 71.0 & 68.0 & \bf 71.0 & 29.0 & \bf 71.0 \\

    cwe-243 & 26.0 & 69.0 & \bf 99.0 & 96.0 & 68.0 & 87.0 & 86.0 & \bf 96.0 \\

    cwe-273 & 63.0 & 61.0 & 16.0 & \bf 77.0 & 51.0 & 38.0 & \bf 92.0 & 73.0 \\

    cwe-311 & \bf 1.0 & 0.0 & 0.0 & 0.0 & 4.0 & 0.0 & \bf 6.0 & 0.0 \\

    cwe-312 & 58.0 & 63.0 & 97.0 & \bf 100.0 & 45.0 & 85.0 & 77.0 & \bf 95.0 \\

    cwe-367 & 75.0 & 95.0 & \bf 97.0 & 92.0 & 94.0 & 76.0 & \bf 96.0 & 94.0 \\

    cwe-369 & 26.0 & 42.0 & \bf 54.0 & 46.0 & \bf 35.0 & 13.0 & 3.0 & 17.0 \\

    cwe-401 & 37.0 & \bf 99.0 & 93.0 & 33.0 & 51.0 & \bf 78.0 & 8.0 & 64.0 \\

    cwe-532 & 3.0 & 0.0 & \bf 50.0 & 16.0 & 13.0 & 1.0 & \bf 19.0 & 5.0 \\

    cwe-643 & 96.0 & 98.0 & 99.0 & \bf 100.0 & 94.0 & 85.0 & \bf 100.0 & \bf 100.0 \\

    cwe-770 & 65.0 & 67.0 & \bf 97.0 & 81.0 & 83.0 & 64.0 & \bf 99.0 & 75.0 \\

    cwe-777 & 4.5 & 7.5 & \bf 20.5 & 20.0 & 3.0 & \bf 32.0 & 27.0 & 14.0 \\

    cwe-788 & 87.0 & 93.0 & \bf 100.0 & \bf 100.0 & 74.0 & 87.0 & \bf 99.0 & 93.0 \\

    cwe-798 & 72.0 & 73.0 & \bf 94.0 & 69.0 & 66.3 & 73.3 & \bf 93.7 & 92.0 \\

    cwe-918 & \bf 47.5 & 10.5 & 17.0 & 14.5 & \bf 24.5 & 8.5 & 2.5 & 1.0 \\

\hline
Total & 47.6 & 60.6 & \bf 69.8 & 64.4 & 51.5 & 60.4 & \bf 66.4 & 61.3 \\

    \bottomrule
    \end{tabular}
    \label{tab:rq1_codellama_results}
}
\end{table*}


\section{Appendix D: Limitations}

LLMs can produce different outputs for the same prompt. To prevent the huge variations, we set a large (100) sample size to ensure the stabilization of results. CWEBench relies on CodeQL to detect vulnerabilities. Such a static analyzer may produce false positives. Although we manually reviewed samples, misclassifications could still influence the conclusions. We performed de-duplication across both our fine-tuning dataset and evaluation benchmarks (CWEBench and CWEval). However, this process may fail to detect duplicates, leading to false negatives.

We follow prior work and choose LLMs with smaller parameter sizes to make fine-tuning feasible under limited computational resources. 
Instruction-tuning larger models, such as ChatGPT-4o, would incur significantly higher costs. As a result, we acknowledge that our findings may not fully generalize to larger LLMs, and we consider this a direction for future investigation. We include 44 distinct CWEs in this study, reflecting a broad range of vulnerabilities. However, numerous additional CWEs remain unaddressed in this study. Moreover, individual CWE can manifest in various real-world scenarios, further highlighting the complexity of comprehensive evaluation. To improve coverage, we expanded our security-sensitive prompts to incorporate additional CWEs beyond those included in earlier studies. Compared to prior work~\citep{pearce2022asleep,he2023large,he2024instruction,li2024exploratory}, our study evaluates a significantly greater number of CWEs, providing a more thorough assessment of generated code vulnerabilities. Future efforts should include broader CWEs.
Additionally, we include six programming languages in our evaluation benchmark. While these languages are highly popular and widely adopted in software development, our findings may not generalize to all programming languages. Future work should expand the language set to strengthen the robustness of the results.
\end{document}